\documentclass[intlimits,twoside,a4paper]{article}

\usepackage{amsmath,amssymb}
\usepackage{graphicx}

\usepackage[T2A]{fontenc}
\usepackage[cp1251]{inputenc}

\usepackage[eqsecnum]{cmpj3}

\issue{2017}{20}{4}{43501}
\doinumber{10.5488/CMP.20.43501}
\title[Isotropic-nematic transition in a HS/HSC mixture]{Isotropic-nematic transition in a mixture of hard spheres and hard spherocylinders: scaled particle theory description}
\author[M.F. Holovko, M.V. Hvozd]{M.F. Holovko, M.V. Hvozd}
\address{Institute for Condensed Matter Physics of the National Academy of Sciences of Ukraine, \\1 Svientsitskii St., 79011 Lviv, Ukraine 
}

\date{Received July 27, 2017}
\begin{document}

\maketitle

\begin{abstract}
The scaled particle theory is developed for the description of thermodynamical properties of a mixture of hard spheres and hard spherocylinders.  Analytical expressions for  free energy,  pressure and chemical potentials are derived. From the minimization of  free energy, a nonlinear integral equation for the orientational singlet distribution function is formulated.  An isotropic-nematic phase transition in this mixture is investigated from the bifurcation analysis of this equation. It is shown that with an increase of  concentration of hard spheres, the total packing fraction of a mixture on phase boundaries slightly increases. The obtained results are compared with computer simulations data.

\pacs 51.30.+i, 64.70.Md
\end{abstract}

\section{Introduction}

A hard spherocylinder fluid is  one of the simplest models widely used for the description of an isotropic-nematic phase transition in liquid crystals \cite{Vroege1992,Franco-Melgar2008}. The first treatment of isotropic-nematic transition in a hard-spherocylinder fluid was performed by Onsager about seventy years ago~\cite{Onsager1949}. The Onsager theory can be considered as the density functional theory in which a low-density expansion of the free energy functional is truncated at the level of second virial coefficient. The equilibrium state is determined by the functional variation of free energy with respect to the orientational distribution function. It was shown \cite{Onsager1949} that such a treatment is exact in the limit of infinitely thin rods when $L\rightarrow\infty$ and $D\rightarrow0$, but $L^2 D$ is fixed, where $L$ and $D$ are the length and the diameter of spherocylinders, respectively. It was shown that besides an isotropic-nematic transition, the Onsager theory describes a nematic-smectic transition in a hard-spherocylinder fluid, which appears at higher densities \cite{Koda1996}. The application of the scaled particle theory (SPT)~\cite{Cotter1970,Cotter1974,Lasher1970,Cotter1979,Cotter1978} provides an efficient approximate way to incorporate the higher-order contributions neglected in the Onsager theory. An alternative way of improving  the Onsager theory is the Parsons-Lee (PL) approach \cite{Parsons1979,Lee1987,Lee1988}, which is based on the mapping of the properties of a spherocylinder fluid to those of the hard-sphere model. The SPT theory was also extended for the description of
a hard-spherocylinder fluid in random porous media \cite{Holovko2014,Holovko2015}.

During the last decades the approaches developed for a hard-spherocylinder fluid have been generalized for the description of mixtures of hard anisotropic particles. In such systems,  new phases were observed and their properties were richer and more complicated than those for the one-component case
\cite{Koda1996,Galindo2003,Cinacchi2004,Lago2004,Vesely2005,Martinez-Raton2006,Cuetos2007,Cuetos2008,Malijevsky2008,Belli2012,Gamez2013,Wu2015}. The simplest example of such multi-component systems of hard anisotropic particles is a binary mixture of hard spheres and hard spherocylinders, for the description of which the corresponding approaches have been proposed.
Among them there are the Onsager theory \cite{Lee1988,Cinacchi2004,Vesely2005,Agren1975} and the Parsons-Lee approach \cite{Malijevsky2008,Gamez2013,Wu2015} in the one-fluid and many-fluid approximations for a hard-spheres mixture and computer simulations \cite{Lago2004,Cuetos2007,Malijevsky2008,Gamez2013,Wu2015}.

In this paper we present a development of the scaled particle theory for the description of a binary mixture of hard spheres and hard spherocylinders. We derive expressions for the chemical potentials of hard sphere and hard spherocylinder components. From the minimization of free energy, a non-linear integral equation for the orientational distribution function is obtained. From the bifurcation analysis of this integral equation, an isotropic-nematic phase diagram of a binary mixture of hard spheres and hard spherocylinders is analysed and discussed. The results of the presented approach are numerically compared  with some computer simulation data.

The paper is arranged as follows. The theoretical part is presented in section~\ref{sec2}. The discussion of the obtained results and the comparison to computer simulations data are given in section~\ref{sec3}.  Finally, we draw some conclusions in the last section.


\section{Theory}\label{sec2}

We consider a two-component hard convex body (HCB) fluid consisting of hard spheres (HS) and hard
spherocylinders (HSC). In order to characterize HCB particles we use three functionals: the volume~$V$, the surface area $S$ and the mean curvature $r$ taken with the factor ${1}/{4\piup}$. For  hard spheres of the radius~$R_{1}$, these functionals are as follows:
\begin{equation}
\label{funct1}
V_{1}=\frac{4}{3}\piup R_{1}^3\,,\qquad   S_{1}=4\piup R_{1}^2\,,\qquad   r_{1}=R_{1}
\end{equation}
and for the hard spherocylinders of the radius $R_{2}$ and of the length $L_{2}$
\begin{equation}
\label{funct2}
V_{2}=\piup R_{2}^2 L_{2}+\frac{4}{3}\piup R_{2}^3\,,\qquad   S_{2}=2\piup R_{2} L_{2}+4\piup R_{2}^2\,,\qquad   r_{2}=\frac{1}{4} L_{2}+R_{2}.
\end{equation}

A basic idea of the SPT approach is an insertion of an additional particle of a variable size, e.g., a scaled
particle, into a fluid.
Adding a scaled hard-sphere particle into our system we use the scaling
parameter $\lambda_{\text s}$. Therefore, the volume $V_{\text{1s}}$, the surface area $S_{\text{1s}}$ and the mean curvature $r_{\text{1s}}$ are modified according to the following relations
\begin{equation}
\label{funct1s}
V_{\text{1s}}=\lambda_{\text s}^3 V_1\,,\qquad S_{\text{1s}}=\lambda_{\text s}^2 S_1\,,\qquad r_{\text{1s}}=\lambda_{\text s} r_1.
\end{equation}
When we add a scaled hard-spherocylinder particle into a fluid with the scaling radius $R_{\text{2s}}$ and the scaling length $L_{\text{2s}}$, in addition to the scaling parameter $\lambda_{\text s}$, we introduce the scaling parameter $\alpha_{\text s}$. Therefore, the radius  $R_{\text{2s}}$ and the length $L_{\text{2s}}$ are defined as \cite{Cotter1970,Cotter1974}
\begin{equation}
\label{funct2s}
R_{\text{2s}}=\lambda_{\text s} R_2\,,\qquad L_{\text{2s}}=\alpha_{\text s} L_2.
\end{equation}
As a result, the volume $V_{\text{2s}}$, the surface area $S_{\text{2s}}$ and the mean curvature $r_{\text{2s}}$ of the scaled spherocylinder are equal to
\begin{equation}
\label{funct2ss}
V_{\text{2s}}=\piup R_2^2 L_2 \alpha_{\text s}\lambda_{\text s}^2+\frac{4}{3} \piup R_2^3 \lambda_{\text s}^3\,,\quad
S_{\text{2s}}=2 \piup R_2 L_2\alpha_{\text s}\lambda_{\text s}+4\piup R_2^2 \lambda_{\text s}^2\,,\quad
r_{\text{2s}}=\frac{1}{4}L_2\alpha_{\text s}+R_2\lambda_{\text s}.
\end{equation}

The procedure of insertion of a scaled particle into a HCB fluid is equivalent to the creation of a corresponding cavity. This cavity is free of centers of any other fluid particles. The key point of the SPT theory is a consideration of the excess chemical potential of a scaled particle $\mu_{\text s}^{\text{ex}}$
\cite{Holovko2009,Chen2010,Patsahan2011,Reiss1959,Reiss1960,Lebowitz1965,Holovko2010}. The work needed to create such a cavity is equal to $\mu_{\text s}^{\text{ex}}$.

For a small scaled HS and HSC particles, the excess chemical potentials can be written in the following form \cite{Patsahan2011}
\begin{eqnarray}
\label{chem1ssmall1}
\beta\mu_{\text{1s}}^{\text{ex}} (\lambda_{\text s})=-\ln \bigg[1-\eta_1(1+\lambda_{\text s})^3
-\eta_2\bigg(1+\frac{r_{\text{1s}}S_2}{V_2}+\frac{r_2 S_{\text{1s}}}{V_2}+\frac{V_{\text{1s}}}{V_2}\bigg)\bigg],
\end{eqnarray}
\begin{eqnarray}
\label{chem2ssmall1}
\beta\mu_{\text{2s}}^{\text{ex}}(\alpha_{\text s},\lambda_{\text s})=
-\ln \bigg[1-\eta_1\left(1+\frac{r_{\text{2s}}S_1}{V_1}+\frac{r_1 S_{\text{2s}}}{V_1}+\frac{V_{\text{2s}}}{V_1}\right)
-\eta_2\bigg(1+\frac{r_{\text{2s}}S_2}{V_2}+\frac{r_2 S_{\text{2s}}}{V_2}+\frac{V_{\text{2s}}}{V_2}\bigg)\bigg],
\end{eqnarray}
where $\beta=1/k_{\text B} T$, $k_{\text B}$ is the Boltzmann constant, $T$ is the temperature, $\eta_1=\rho_1 V_1$ is the HS fluid packing fraction, $\rho_1$ is the HS fluid density, $V_1$ is the volume of a HS particle;  $\eta_2=\rho_2 V_2$ is the HSC fluid packing fraction, $\rho_2$ is the HSC fluid density, $V_2$ is the volume of a HSC particle.
It should be noted that equation~(\ref{chem2ssmall1}) is written for an isotropic case.

After  substituting equations~(\ref{funct1})--(\ref{funct2ss}) into equations~(\ref{chem1ssmall1})--(\ref{chem2ssmall1}) and having generalized  equation~(\ref{chem2ssmall1}) for the anisotropic case, the chemical potentials of the HS and HSC scaled particles can be presented as follows:
\begin{align}
\label{chem1ssmall}
&\beta\mu_{\text{1s}}^{\text{ex}} (\lambda_{\text s})=-\ln \bigg\{1-\eta_1(1+\lambda_{\text s})^3 -\eta_2 \bigg[1+\frac{1}{ k_1}\frac{6\gamma_2}{3\gamma_2-1}
\lambda_{\text s}+\frac{1}{k_1^2}\frac{3(\gamma_2+1)}{3\gamma_2-1}
\lambda_{\text s}^2+\frac{1}{k_1^3}\frac{2}{3\gamma_2-1} \lambda_{\text s}^3 \bigg]\bigg\},
\\
\label{chem2ssmall}
&\beta\mu_{\text{2s}}^{\text{ex}}(\alpha_{\text s},\lambda_{\text s})=-\ln \bigg(1-\eta_1\left[\frac{3}{4} s_1\alpha_{\text s}\left(1+k_1\lambda_{\text s}\right)^2+\left(1+k_1\lambda_{\text s}\right)^3\right] \nonumber \\
&-\eta_2\bigg\{1+\frac{3(\gamma_2-1)}{3\gamma_2-1}\left[1+(\gamma_2-1)\tau(f)\right]\alpha_{\text s}
+\frac{6\gamma_2}{3\gamma_2-1}\lambda_{\text s}
+\frac{6(\gamma_2-1)}{3\gamma_2-1}\left[1+\frac{1}{2}(\gamma_2-1)\tau(f)\right]\alpha_{\text s}\lambda_{\text s}
\nonumber\\
&+\frac{3(\gamma_2+1)}{3\gamma_2-1}\lambda_{\text s}^2+\frac{3(\gamma_2-1)}{3\gamma_2-1}\alpha_{\text s}\lambda_{\text s}^2
+\frac{2}{3\gamma_2-1}\lambda_{\text s}^3\bigg\}\bigg),
\end{align}
where
$k_1$, $s_1$ and $\gamma_2$ are equal to
\begin{eqnarray}
\label{k1s1gamma2}
k_1=\frac{R_2}{R_1}\,,\qquad s_1=\frac{L_2}{R_1}\,,\qquad \gamma_2=1+\frac{L_2}{2 R_2}\,,
\end{eqnarray}
and
\begin{equation}
\tau(f)=\frac{4}{\piup}\int f(\Omega_1)f(\Omega_2)\sin\gamma(\Omega_1,\Omega_2)\rd\Omega_1\rd\Omega_2.
\end{equation}
Here, $\Omega=(\vartheta,\varphi)$ denotes the orientation of HSC particles and it is defined by the angles $\vartheta$ and $\varphi$, where $\rd\Omega=\frac{1}{4\piup}\sin\vartheta \rd\vartheta \rd\varphi$ is the normalized angle element, $\gamma(\Omega_1, \Omega_2)$ is the angle between orientational vectors of two molecules, $f(\Omega)$ is the singlet orientation distribution function normalized in such a way that
\begin{equation}
\label{normalization}
\int f(\Omega)\rd\Omega=1.
\end{equation}
$f(\Omega)$ is defined herein below from the minimization of the free energy of a considered mixture.

For a large scaled particle, the excess chemical potential is equal to the work needed to create a macroscopic cavity within the fluid and it is given by a thermodynamic expression. For a scaled hard sphere particle, it can be presented as follows:
\begin{equation}
\label{chem1slarge}
\beta\mu_{\text{1s}}^{\text{ex}}=w(\lambda_{\text s})+\beta PV_{\text{1s}}\,,
\end{equation}
where $P$ is the pressure of a fluid and $V_{\text{1s}}$ is the volume of a scaled HS particle.
Similarly, for a scaled HSC particle, we have
\begin{equation}
\label{chem2slarge}
\beta\mu_{\text{2s}}^{\text{ex}}=w(\alpha_{\text s},\lambda_{\text s})+\beta PV_{\text{2s}}\,,
\end{equation}
where $P$ and $V_{\text{2s}}$ are the pressure of a fluid and the volume of a scaled HSC particle, respectively.

According to the ansatz of SPT
\cite{Holovko2015,Holovko2009,Chen2010,Patsahan2011,Reiss1959,Reiss1960,Lebowitz1965,Holovko2010}
 $w(\lambda_{\text s})$ and $w(\alpha_{\text s},\lambda_{\text s})$
can be written in the form of expansions:
\begin{eqnarray}
\label{expansion1s}
&& w(\lambda_{\text s})=w_0+w_1\lambda_{\text s}+\frac{1}{2} w_2\lambda_{\text s}^2\,, \\
\label{expansion2s}
&& w(\alpha_{\text s},\lambda_{\text s})=w_{00}+w_{01}\alpha_{\text s}+w_{10}\lambda_{\text s}+w_{11}\alpha_{\text s}\lambda_{\text s}+\frac{1}{2} w_{20}\lambda_{\text s}^2.
\end{eqnarray}
We can derive the coefficients of these expansions from the continuity of $\mu_{\text{1s}}^{\text{ex}}$, $\mu_{\text{2s}}^{\text{ex}}$ and their corresponding derivatives ${\partial\mu_{\text{1s}}^{\text{ex}}}/{\partial\lambda_{\text s}}$,  ${\partial^2\mu_{\text{1s}}^{\text{ex}}}/{\partial\lambda_{\text s}^2}$ at $\lambda_{\text s}=0$ for a scaled HS particle; ${\partial\mu_{\text{2s}}^{\text{ex}}}/{\partial\alpha_{\text s}}$, ${\partial\mu_{\text{2s}}^{\text{ex}}}/{\partial\lambda_{\text s}}$, ${\partial^2\mu_{\text{2s}}^{\text{ex}}}/{\partial\alpha_{\text s}\partial\lambda_{\text s}}$, ${\partial^2\mu_{\text{2s}}^{\text{ex}}}/{\partial\lambda_{\text s}^2}$ at $\alpha_{\text s}=\lambda_{\text s}=0$ for a scaled HSC particle.

Therefore, for a scaled HS particle, we obtain:
\begin{eqnarray}
\label{coef1s0}
&& w_0=-\ln(1-\eta)\nonumber, \\
\label{coef1s1}
&& w_1=3\frac{\eta_1}{1-\eta}+\frac{1}{k_1}\frac{6\gamma_2}{3\gamma_2-1}\frac{\eta_2}{1-\eta}\,, \nonumber\\
\label{coef1s2}
&& w_2=6\frac{\eta_1}{1-\eta}+\frac{1}{k_1^2}\frac{6(\gamma_2+1)}{3\gamma_2-1}\frac{\eta_2}{1-\eta}
+\frac{1}{(1-\eta)^2}\bigg(3\eta_1+\frac{1}{k_{1}}\frac{6\gamma_2}{3\gamma_2-1}\eta_2\bigg)^2,
\end{eqnarray}
where $\eta=\eta_1+\eta_2$ is the total packing fraction of the mixture considered.
For a scaled HSC particle, we find:
\begin{align}
& w_{00}=-\ln(1-\eta)\nonumber, \\
& w_{01}=\frac{3}{4}s_1\frac{\eta_1}{1-\eta}
+\left[\frac{3(\gamma_2-1)}{3\gamma_2-1}+\frac{3(\gamma_2-1)^2\tau(f)}{3\gamma_2-1}\right]\frac{\eta_2}{1-\eta}
\nonumber\,, \\
& w_{10}=3k_1\frac{\eta_1}{1-\eta}+\frac{6\gamma_2}{3\gamma_1-1}\frac{\eta_2}{1-\eta}\nonumber\,, \\
& w_{11}=\frac{3}{2}k_1s_1\frac{\eta_1}{1-\eta}
+\left[\frac{6(\gamma_2-1)}{3\gamma_2-1}+\frac{3(\gamma_2-1)^2\tau(f)}{3\gamma_2-1}\right]\frac{\eta_2}{1-\eta}\nonumber\\
& +\left\{\frac{3}{4}s_1\frac{\eta_1}{1-\eta}
+\left[\frac{3(\gamma_2-1)}{3\gamma_2-1}+\frac{3(\gamma_2-1)^2\tau(f)}{3\gamma_2-1}\right]\frac{\eta_2}{1-\eta}\right\}
\left(3k_1\frac{\eta_1}{1-\eta}+\frac{6\gamma_2}{3\gamma_2-1}\frac{\eta_2}{1-\eta}\right)\nonumber, \\
\label{coef2s20}
&  w_{20}=6k_1^2\frac{\eta_1}{1-\eta}+\frac{6(\gamma_2+1)}{3\gamma_1-1}\frac{\eta_2}{1-\eta}
+\frac{1}{(1-\eta)^2}\left(3k_1\eta_1+\frac{6\gamma_2}{3\gamma_2-1}\eta_2\right)^2.
\end{align}

After setting $\lambda_{\text s}=1$ in equation~(\ref{expansion1s}) and $\alpha_{\text s}=\lambda_{\text s}=1$ in equation~(\ref{expansion2s}), the HS and HSC scaled particles become of the same sizes as HS and HSC particles of a fluid, respectively. It makes it possible to find the relation between the pressure and the excess chemical potentials $\mu_{1}^{\text{ex}}$ and  $\mu_{2}^{\text{ex}}$ of a fluid. The total chemical potentials for HS and HSC particles in a HS/HSC mixture are as follows:
\begin{eqnarray}
\label{chem1and2total}
&& \beta\mu_{1}=\ln(\rho_1\Lambda_1^3)+\beta\mu_{1}^{\text{ex}}, \\
&& \beta\mu_{2}=\ln(\rho_2\Lambda_2^3\Lambda_{\text{2R}})+\beta\mu_{2}^{\text{ex}},
\end{eqnarray}
where $\Lambda_{1}$ and $\Lambda_{2}$ are the fluid thermal wavelengths of the HS and HSC components, respectively; $\Lambda_{\text{2R}}^{-1}$ is the rotational partition function of a single HSC molecule \cite{Gray1984}.
Then, we can write expressions for the total chemical potentials as  follows:
\begin{eqnarray}
\label{chem1total}
\beta\mu_{1}=\ln(\rho_1\Lambda_1^3)-\ln(1-\eta)+a_1\frac{\eta}{1-\eta}
+b_1\frac{\eta^2}{(1-\eta)^2}+\beta P\frac{\eta_1}{\rho_1}\,,
\end{eqnarray}
\begin{eqnarray}
\label{chem2total}
\beta\mu_{2}=\ln(\rho_2\Lambda_2^3\Lambda_{\text{2R}})-\ln(1-\eta)+a_2\frac{\eta}{1-\eta}
+b_2\frac{\eta^2}{(1-\eta)^2}+\beta P\frac{\eta_2}{\rho_2}\,,
\end{eqnarray}
where the coefficients $a_1$,  $a_2$, $b_1$ and $b_2$ are:
\begin{eqnarray}
\label{a1}
a_1=6\frac{\eta_1}{\eta}+\left[\frac{1}{k_1}\frac{6\gamma_2}{3\gamma_2-1}
+\frac{1}{k_1^2}\frac{3(\gamma_2+1)}{3\gamma_2-1}\right]\frac{\eta_2}{\eta}\nonumber\,,
\end{eqnarray}
\begin{eqnarray}
\label{b1}
b_1=\frac{1}{2}\left(3\frac{\eta_1}{\eta}+\frac{1}{k_1}\frac{6\gamma_2}{3\gamma_2-1}\frac{\eta_2}{\eta}\right)^2
\end{eqnarray}
and
\begin{eqnarray}
\label{a2}
a_2\big(\tau(f)\big)=\left[\frac{3}{4}s_1(1+2k_1)+3k_1(1+k_1)\right]\frac{\eta_1}{\eta}
+\left[6+\frac{6(\gamma_2-1)^2\tau(f)}{3\gamma_2-1}\right]\frac{\eta_2}{\eta}\,, \nonumber
\end{eqnarray}
\vspace{-6mm}
\begin{eqnarray}
\label{b2}
b_2\big(\tau(f)\big)=\left\{\left(\frac{3}{4}s_1+\frac{3}{2}k_1\right)\frac{\eta_1}{\eta}
+\left[\frac{3(2\gamma_2-1)}{3\gamma_2-1}+\frac{3(\gamma_2-1)^2\tau(f)}{3\gamma_2-1}\right]\frac{\eta_2}{\eta}\right\}
\left(3k_1\frac{\eta_1}{\eta}+\frac{6\gamma_2}{3\gamma_1-1}\frac{\eta_2}{\eta}\right).
\end{eqnarray}
Therefore, we have two equations (\ref{chem1total}) and (\ref{chem2total}), and each of them contains two unknown
quantities: one of the chemical potentials and the pressure. In the case of an one-component fluid, we can eliminate one of these unknowns, $\beta\mu_{1}$ ($\beta\mu_{2}$) or $P$, using equation~(\ref{chem1total}) or equation~(\ref{chem2total}) and the Gibbs-Duhem relation. In our case, the Gibbs-Duhem equation has the form
\begin{equation}
\label{GibbsDuhem1}
\frac{\partial(\beta P)}{\partial\rho}=\sum_{\alpha=1}^{2} \rho_\alpha\frac{\partial(\beta\mu_\alpha)}{\partial\rho}.
\end{equation}

Expressions for the chemical potentials can be obtained according to the recent paper \cite{Chen2016}, in which an expression for the pressure is obtained from equation~(\ref{GibbsDuhem1}), and an expression for the free energy is obtained from an integration over the total density $\rho$. From the differentiation of the free energy with
respect to $\rho_1$ and $\rho_2$, we derive  expressions for the chemical potentials $\mu_1$ and $\mu_2$.

In order to use equation~(\ref{GibbsDuhem1}) and to get one equation containing only one unknown instead of equations~(\ref{chem1total}) and (\ref{chem2total}), we take the derivatives with respect to the total fluid density $\rho=\sum\nolimits_{\alpha=1}^{2}{\rho_\alpha}$ on the both sides of equations~(\ref{chem1total}) and (\ref{chem2total}) keeping the fluid composition constant: $x_\alpha=\rho_\alpha/\rho$, $\alpha=1,2$. Hence, we derive
\begin{eqnarray}
\label{dchem1}
\frac{\partial(\beta\mu_1)}{\partial\rho}=\frac{1}{\rho}\left[1+\frac{\eta}{1-\eta}
+a_1\frac{\eta}{(1-\eta)^2}+2b_1\frac{\eta^2}{(1-\eta)^3}\right]+\frac{4}{3}\piup R_1^3\frac{\partial(\beta P)}{\partial\rho}\,,
\end{eqnarray}
\begin{eqnarray}
\label{dchem2}
\frac{\partial(\beta\mu_2)}{\partial\rho}=\frac{1}{\rho}\left[1+\frac{\eta}{1-\eta}
+a_2\frac{\eta}{(1-\eta)^2}+2b_2\frac{\eta^2}{(1-\eta)^3}\right]
+\left(\piup R_2^2 L_2+\frac{4}{3}\piup R_2^3\right)\frac{\partial(\beta P)}{\partial\rho}.
\end{eqnarray}
The combination of equations~(\ref{GibbsDuhem1}) and (\ref{dchem1})--(\ref{dchem2}) makes it possible to write an expression for the fluid compressibility. Taking into account that $\sum\nolimits_{\alpha}{x_\alpha}=1$, we obtain
\begin{equation}
\label{dpressure}
\frac{\partial(\beta P)}{\partial\rho}=\frac{1}{1-\eta}+(1+A)\frac{\eta}{(1-\eta)^2}
+(A+2B)\frac{\eta^2}{(1-\eta)^3}+2B\frac{\eta^3}{(1-\eta)^4}\,,
\end{equation}
where
\begin{equation}
A=\sum_{\alpha=1}^{2}x_\alpha a_\alpha\,, \qquad
\label{B}
B=\sum_{\alpha=1}^{2}x_\alpha b_\alpha.
\end{equation}
From the integration of equation~(\ref{dpressure}) over the total density $\rho$ at a constant concentration, we find
\begin{equation}
\label{pressure}
\frac{\beta P}{\rho}=1+\frac{\eta}{1-\eta}+\frac{A}{2}\frac{\eta}{(1-\eta)^2}
+\frac{2B}{3}\frac{\eta^2}{(1-\eta)^3}.
\end{equation}

Now, we calculate the Helmholtz free energy, which is related to the pressure as
\begin{equation}
\label{freeenergyint}
\frac{\beta F}{V}=\rho\int_{0}^{\rho} \rd\rho'\frac{1}{\rho'}\left(\frac{\beta P}{\rho'}\right).
\end{equation}
We carry out this integration at  fixed concentrations $x_\alpha$, where $\alpha=1,2$.
Thus, the final expression for the free energy is
\begin{equation}
\label{freeenergy}
\frac{\beta F}{V}=\frac{\beta F_{\text{id}}}{V}+\rho\left[-\ln(1-\eta)+\frac{A}{2}\frac{\eta}{1-\eta}+\frac{B}{3}\frac{\eta^2}{(1-\eta)^2}\right],
\end{equation}
where  $F_{\text{id}}$ is the ideal gas contribution to the Helmholtz free energy of a mixture:
\begin{equation}
\label{Fid}
\frac{\beta F_{\text{id}}}{V}=\rho_1\left[\ln(\Lambda_1^3\rho_1)-1\right]
+\rho_2\left[\ln(\Lambda_2^3\rho_2)-1\right]+\rho_2\sigma(f).
\end{equation}
Here, $\sigma(f)$ is the entropic term defined as
\begin{equation}
\label{sigma}
\sigma(f)=\int f(\Omega)\ln f(\Omega)\rd\Omega.
\end{equation}

The singlet orientational distribution function $f(\Omega)$ can be obtained from the minimization of free energy with respect to variations of this distribution. This procedure leads to a nonlinear integral equation
\begin{equation}
\label{nonlineareq}
\ln f(\Omega_1)+\lambda+C\int f(\Omega')\sin\gamma(\Omega_1\Omega')\rd\Omega'=0,
\end{equation}
where
\begin{equation}
\label{constC}
C=\frac{\eta_2}{1-\eta}\left[\frac{3(\gamma_2-1)^2}{3\gamma_2-1}+
\frac{1}{1-\eta}\frac{(\gamma_2-1)^2}{3\gamma_2-1}
\left(3k_1\eta_1+\frac{6\gamma_2}{3\gamma_2-1}\eta_2\right)\right].
\end{equation}
The constant $\lambda$ is defined from the normalization condition equation~(\ref{normalization}).

Using the expression for the Helmholtz free energy, we calculate the total chemical potentials for
the components of HS and HSC in a mixture. From the relationship
\begin{equation}
\label{chemalpha}
\beta\mu_{\alpha}=\frac{\partial}{\partial\rho_{\alpha}}\left(\frac{\beta F}{V}\right),
\end{equation}
we derive
\begin{align}
\label{chem1}
& \beta\mu_{1}=\ln\Lambda_1^3\rho_1-\ln(1-\eta)
+\frac{1}{2}\frac{\eta}{1-\eta}\bigg\{a_1+6\frac{\rho_1 V_1}{\eta}+\frac{\rho_2 V_1}{\eta}\left[\frac{3}{4}s_1(1+2k_1)+3k_1(1+k_1)\right]\bigg\}\nonumber \\
& +\frac{1}{3}\frac{\eta^2}{(1-\eta)^2}
\bigg[b_1+3\frac{\rho_1 V_1} {\eta^2}\left(3\eta_1+\frac{1}{k_1}\frac{6\gamma_2}{3\gamma_2-1}\eta_2\right)
\nonumber \\
& +\frac{\rho_2 V_1}{\eta^2}\bigg(9k_1\left(\frac{1}{2}s_1+k_1\right)\eta_1
+\left\{\frac{3}{4}\frac{6\gamma_2}{3\gamma_2-1}s_1
+3k_1\left[3+\frac{3(\gamma_2-1)^2\tau(f)}{3\gamma_2-1}\right]\right\}\eta_2\bigg)\bigg]
+\beta P V_1
\end{align}
for the chemical potential of HS and
\begin{align}
\label{chem2}
& \beta\mu_{2}=\ln\Lambda_2^3\rho_2+\sigma(f)-\ln(1-\eta)
\nonumber \\
& +\frac{1}{2}\frac{\eta}{1-\eta}\bigg\{a_2+\frac{\rho_1 V_2}{\eta}
\left[\frac{1}{k_1}\frac{6\gamma_2}{3\gamma_2-1}
+\frac{1}{2}\frac{1}{k_1^2}\frac{6(\gamma_2+1)}{3\gamma_2-1}\right]
+\frac{\rho_2 V_2}{\eta}\left[6+\frac{6(\gamma_2-1)^2\tau(f)}{3\gamma_2-1}\right]\bigg\}
\nonumber\\
& +\frac{1}{3}\frac{\eta^2}{(1-\eta)^2}
\bigg[b_2+\frac{\rho_1 V_2}{\eta^2}\frac{1}{k_1}\frac{6\gamma_2}{3\gamma_2-1}
\left(3\eta_1+\frac{1}{k_1}\frac{6\gamma_2}{3\gamma_2-1}\eta_2\right)
\nonumber \\
& +\frac{\rho_2 V_2}{\eta^2}\bigg(\left\{\frac{3}{4}\frac{6\gamma_2}{3\gamma_2-1}s_1
+3k_1\left[3+\frac{3(\gamma_2-1)^2\tau(f)}{3\gamma_2-1}\right]\right\}\eta_1 \nonumber \\
& +\frac{6\gamma_2}{3\gamma_2-1}\left[\frac{6(2\gamma_2-1)}{3\gamma_2-1}
+\frac{6(\gamma_2-1)^2\tau(f)}{3\gamma_2-1}\right]\eta_2\bigg)\bigg]
+\beta P V_2
\end{align}
for the chemical potential of HSC.

\section{Results and discussions}\label{sec3}

We use the theory presented in the previous section to study the effect of hard spheres on the isotropic-nematic phase transition in a binary mixture of hard spheres and hard spherocylinders. This investigation is done within
the framework of bifurcation analysis of the integral equation equation~(\ref{nonlineareq}) for the singlet distribution function $f(\Omega)$. It is worth noting that for the first time this equation was obtained by Onsager \cite{Onsager1949} for a hard-spherocylinder fluid in the limit $L_2\rightarrow\infty$ and $R_2\rightarrow0$,
when the dimensionless density of a spherocylinder fluid $c_2=\frac{1}{2}\piup\rho_2 L_2^2 R_2$ was fixed.
Therefore, in the Onsager limit we have
\begin{equation}
\label{constCOnsager}
C\rightarrow c_2=\frac{1}{2}\piup\rho_2 L_2^2 R_2.
\end{equation}

The result, equation~(\ref{constC}), for $C$ is the generalization of the SPT result for a HSC fluid for
the finite values of $L_2$ and $R_2$ \cite{Cotter1978,Tuinier2007}. In this case,
\begin{equation}
\label{constCTuinier2007}
C\rightarrow \frac{\eta_2}{1-\eta_2}\bigg[\frac{3\left(\gamma_2-1\right)^2}{3\gamma_2-1}+
\frac{\eta_2}{1-\eta_2}\frac{6\gamma_2\left(\gamma_2-1\right)^2}{\left(3\gamma_2-1\right)^2}\bigg].
\end{equation}
From the bifurcation analysis of the integral equation equation~(\ref{nonlineareq}) for the singlet distribution function $f(\Omega)$, it was found that this equation has two characteristic points $C_{\text i}$ and $C_{\text n}$ \cite{Kayser1978}, which defined the range of stability of a considered mixture. The first point $C_{\text i}$ corresponds to the highest possible density of the stable isotropic state and the second point $C_{\text n}$ corresponds the lowest possible density of a stable nematic state. For the Onsager model, from the minimization of the free energy with respect to the singlet distribution function $f(\Omega)$, and subsequently from the solution of the coexistence equations,
the following values of density of coexisting isotropic and nematic phases were obtained \cite{Herzfeld1984,Lekkerkerker1984,Chen1993}:
\begin{equation}
\label{ciandcn}
c_{\text i}=3.289,\qquad c_{\text n}=4.192.
\end{equation}

In the presence of hard spheres for the Onsager model, we have
\begin{equation}
\label{COnzLimitHS}
C=\frac{c_2}{1-\eta_1}.
\end{equation}
It means that the isotropic-nematic transition in the presence of hard spheres shifts to lower densities of spherocylinders.

For the binary mixture of hard spheres and hard spherocylinders at the finite value of $L_2/2 R_2$, we can put
\begin{equation}
\label{CiandCn}
C_{\text i}=3.289,\qquad C_{\text n}=4.192,
\end{equation}
where $C_{\text i}$ and $C_{\text n}$ are determined from equation~(\ref{constC}). The values of $C_{\text i}$ and $C_{\text n}$ in equation~(\ref{CiandCn}) define the isotropic-nematic phase diagram for a HS/HSC mixture depending on the ratios $L_2/R_2=2\left(\gamma_2-1\right)$ and $k_1=R_2/R_1$, as well as on the densities of HS and HSC particles, $\eta_1$ and $\eta_2$, respectively. We note that $s_1$ defined by equation~(\ref{k1s1gamma2}) is not an independent parameter, since
\begin{equation}
\label{s1discussion}
s_1=2\left(\gamma_2-1\right) k_1.
\end{equation}

\begin{figure}[!t]
\centerline{
\includegraphics[width=0.46\textwidth]{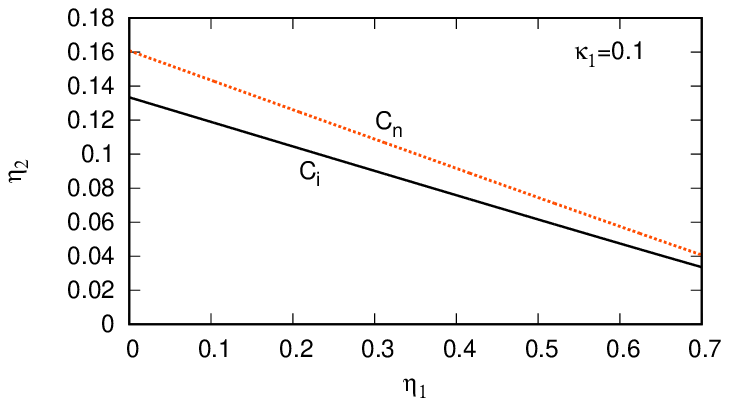}
\includegraphics[width=0.46\textwidth]{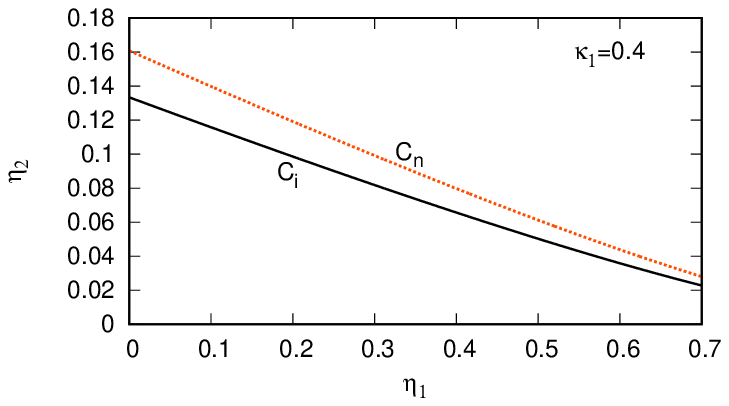}}
\centerline{
\includegraphics[width=0.46\textwidth]{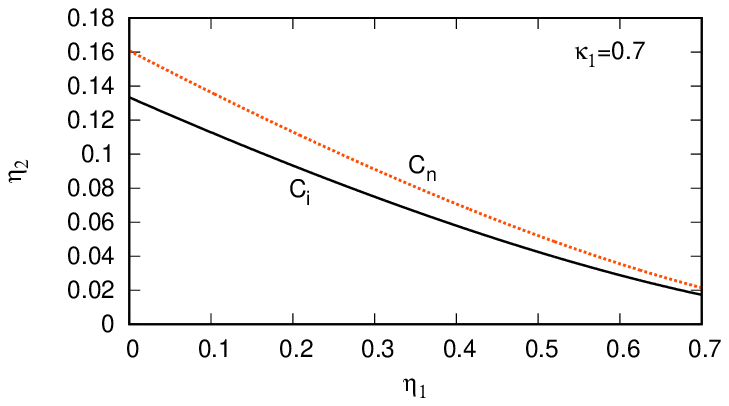}
\includegraphics[width=0.46\textwidth]{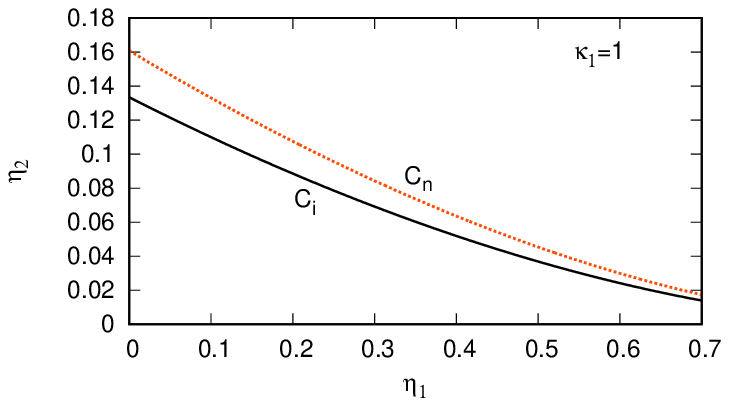}}
\caption{(Color online) Coexistence lines of isotropic-nematic phases of a HS/HSC mixture
for $L_2/2 R_2=20$ presented as a dependence of the packing fraction of HSC particles
$\eta_2=\rho_2 V_2$ on the packing fraction of HS particles $\eta_1=\rho_1 V_1$ at fixed $k_1=R_2/R_1$.
The black line below denoted by $C_{\text i}$ corresponds to the isotropic phase, the red line above
denoted by $C_{\text n}$ corresponds to the nematic phase. The area between the solid black and dotted red lines corresponds to the region of the coexistence of isotropic and nematic phases.}
\label{k1_fixed}
\end{figure}

The packing fraction of hard spheres $\eta_1$ as a function of the packing fraction of hard spherocylinders~$\eta_2$
for $\gamma_2=21$ along the boundaries of isotropic-nematic phase transition is estimated from equation~(\ref{CiandCn})
for a HS/HSC mixture at a fixed ratio $k_1$. As it is seen from figure~\ref{k1_fixed}, the presence of hard spheres shifts the phase transition to  lower densities of hard spherocylinders. Moreover, the interfacial region becomes broader if the size of hard spheres increases ($k_1$ decreases).

\begin{figure}[!t]
\centerline{
\includegraphics[width=0.46\textwidth]{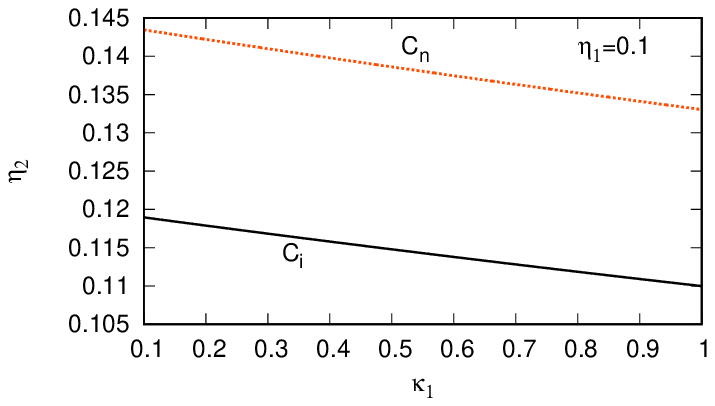}
\includegraphics[width=0.46\textwidth]{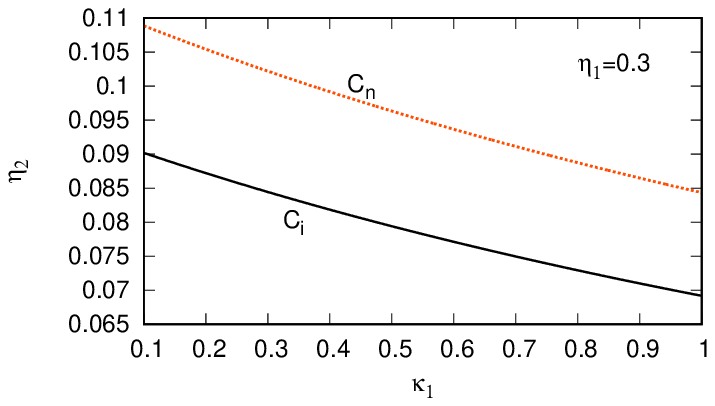}}
\centerline{
\includegraphics[width=0.46\textwidth]{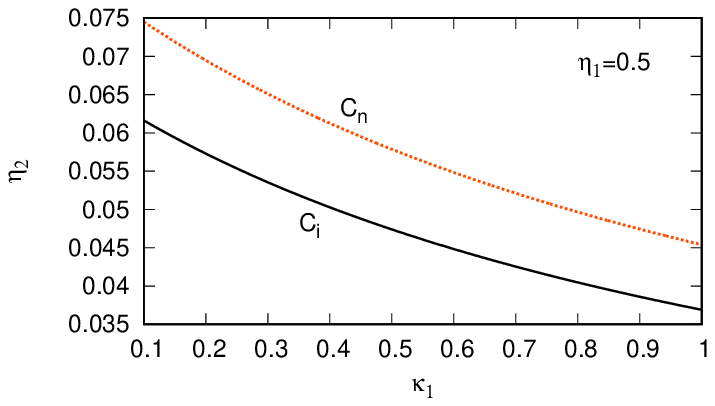}
\includegraphics[width=0.46\textwidth]{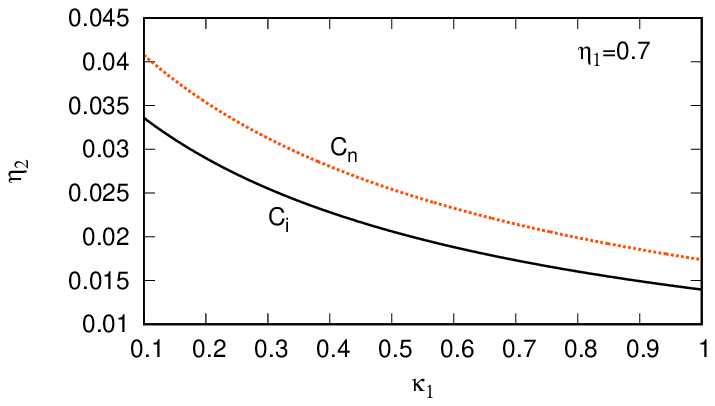}}
\caption{(Color online) Coexistence lines of isotropic-nematic phases of a HS/HSC mixture
for $L_2/2 R_2=20$ presented as a dependence of the packing fraction of HSC particles
$\eta_2=\rho_2 V_2$ on the ratio $k_1=R_2/R_1$ at the fixed packing fraction of HS particles $\eta_1=\rho_1 V_1$.
The black line below denoted by $C_{\text i}$ corresponds to the isotropic phase, the red line above denoted by $C_{\text n}$
corresponds to the nematic phase. The area between the solid black and dotted red lines corresponds to
the region of the coexistence of isotropic and nematic phases.}
\label{eta1_fixed}
\end{figure}

The effect of the size of hard spheres on the isotropic-nematic phase boundaries of the same HS/HCS mixture, but
at a fixed packing fraction $\eta_1$, is demonstrated in figure~\ref{eta1_fixed}. One can observe that
an increase of the packing fraction of hard spheres leads to a contraction of interfacial region.

\begin{figure}[!t]
\centerline{
\includegraphics[width=0.6\textwidth]{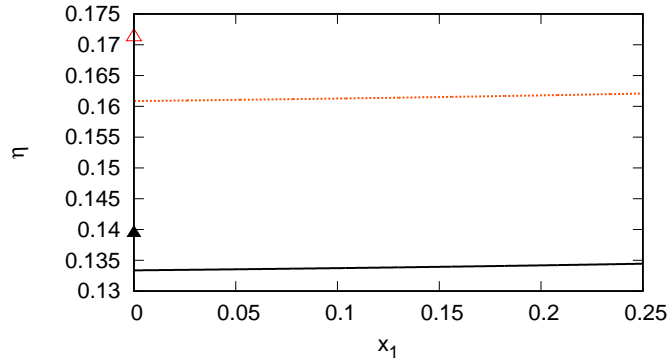}}
\caption{(Color online) Coexistence lines of isotropic-nematic phases of a HS/HSC mixture
for $L_2/2 R_2=20$ and $k_1=1$ presented as a dependence of the packing fraction of HS/HSC mixture $\eta=\eta_1+\eta_2$ on the composition of spherical particles $x_1=\rho_1/(\rho_1+\rho_2)$. For the case of pure HSC system ($x_1=0$) the SPT results are close to those obtained in computer 
 simulations with a use of the modified Gibbs-Duhem integration procedure (filled black triangle for the isotropic branch and open red triangle for the nematic branch) \cite{Bolhuis1997}.}
\label{frenkel}
\end{figure}
\begin{figure}[!t]
\centerline{
\includegraphics[width=0.5\textwidth]{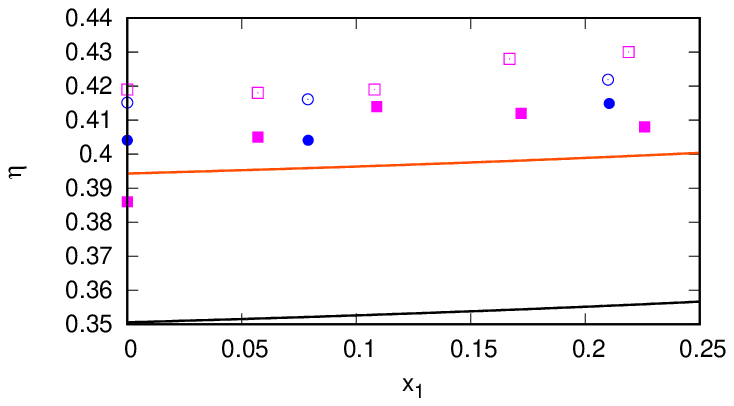}
\includegraphics[width=0.5\textwidth]{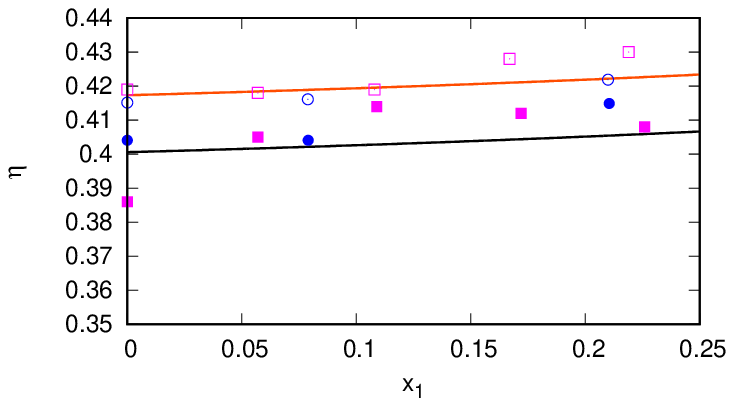}}
\caption{(Color online) Coexistence lines of the isotropic and nematic phases of a HS/HSC mixture
for $L_2/2 R_2=5$ and $k_1=1$ presented as a dependence of the packing fraction of HS/HSC mixture $\eta=\eta_1+\eta_2$ on the composition of spherical particles $x_1=\rho_1/(\rho_1+\rho_2)$. The computer simulations data taken from \cite{Lago2004} are denoted by filled circles for the isotropic branch and by open circles for the nematic branch. The computer simulations data taken from \cite{Wu2015} are shown as filled squares for the isotropic branch and open squares for the nematic branch.}
\label{wuandwured}
\end{figure}

The boundaries of an isotropic-nematic phase transition for the same model are presented in figure~\ref{frenkel} using
the coordinates $\eta=\eta_1+\eta_2$ and $x_1=\rho_1/(\rho_1+\rho_2)$. It is seen that an increase of the composition of spherical particles $x_1$ makes the total packing fraction $\eta$ slightly higher in comparison with the results
obtained in the coordinates $\eta_1$ and $\eta_2$ (see figure~\ref{k1_fixed}). In the case of a pure HSC fluid ($x_1=0$), the SPT results are rather close to the computer simulations data taken from \cite{Bolhuis1997}.
It is noticed that the SPT theory underestimates the values of $\eta_{\text i}$ and $\eta_{\text n}$, and the difference from
the computer simulations increases with a decrease of the parameter $L_2/2 R_2$.

The isotropic-nematic phase transition boundaries for a HS/HSC mixture when
$\gamma_2=6$ and $k_1=1$ are presented in figure~\ref{wuandwured} (the left-hand panel) using the coordinates $\eta$ and $x_1$. A comparison with the computer simulations data taken from \cite{Lago2004} and \cite{Wu2015} are also shown in this figure. It is found that the theoretical prediction of the isotropic line is about $\Delta\eta=0.05$ lower than the simulations results, while for the nematic line it is approximately $\Delta\eta=0.023$ lower. On the other hand,
qualitatively the effect of the composition $x_1$ predicted by the theory and the one obtained from the computer simulations are similar. Hence, in figure~\ref{wuandwured} (the right-hand panel) we present the modified results of a theoretical prediction, where the coexistence lines are shifted up by $\Delta\eta=0.05$ for the isotropic phase and by $\Delta\eta=0.023$ for the nematic phase. As one can see, in this case an agreement between  theoretical and simulation results is rather satisfactory.

It is worth noting that the isotropic-nematic coexistence lines can be also obtained from the condition of thermodynamic equilibrium. According to this condition, the both phases should have the same pressure and the same chemical potentials:
\begin{equation}
\label{coexistence}
P_{\text i}(\eta_{\text i},x_{\text i})=P_{\text n}(\eta_{\text n},x_{\text n}),\qquad \mu_{i,1}(\eta_{\text i},x_{\text i})=\mu_{{\text n},1}(\eta_{\text n},x_{\text n}),
\qquad \mu_{i,2}(\eta_{\text i},x_{\text i})=\mu_{{\text n},2}(\eta_{\text n},x_{\text n}),
\end{equation}
where $\mu_{{\text i},1}$ (or $\mu_{{\text i},2}$) and $\mu_{{\text n},1}$ (or $\mu_{{\text n},2}$) are the chemical potentials of HS (or HSC) particles in the isotropic and nematic phases, respectively.

In \cite{Kayser1978} it was shown that for the Onsager model the results obtained from the bifurcation analysis and from the thermodynamic calculations coincide exactly. We also observe the same for the mixture of Onsager spherocylinders and hard spheres. On the other hand, it was found in \cite{Holovko2015} that for the finite values of $L_2/2 R_2$, there is some difference between the results obtained from these two different approaches, and the difference slightly increases with a decrease of the ratio $L_2/2 R_2$.

\section{Conclusions}
\looseness=-1 In this paper we have generalized the scaled particle theory for the investigation of thermodynamic properties of a mixture of hard spheres and hard spherocylinders. The expressions for the chemical potentials of hard spheres and hard spherocylinders are derived from the consideration of a scaled hard sphere and a scaled hard spherocylinder inserted into a system under study. Analytical expressions for the free energy and for the pressure of the considered mixture are also obtained. From the minimization of the free energy, a nonlinear integral equation for the orientational distribution function is obtained. From the bifurcation analysis of this integral equation, an isotropic-nematic phase transition in a mixture of hard spheres and hard spherocylinders is investigated. It is shown that the presence of hard spheres shifts the phase transition to the lower densities of hard spherocylinders. With an increase of the sizes of hard spheres, the interfacial region is expanded and with an increase of the packing fraction of hard spheres, the interfacial region decreases. It is also shown that with an increase of concentration of hard spheres, the total packing fraction of a mixture on the phase boundaries slightly increases in comparison with phase boundaries in the coordinates of the packing fraction of hard spheres $\eta_1$ and the packing fraction of hard spherocylinders. The obtained results are qualitatively in agreement with computer simulations data.

The present work can be extended directly to the presence of disordered porous media. For the present time, the scaled particle theory for a hard sphere fluid in disordered porous media is quite well developed \cite{Holovko2009,Chen2010,Patsahan2011,Holovko2010,HolPat12,Holovko2013} and has found applications
in describing a reference system within the perturbation theory for fluids with
different types of attraction, such as associative \cite{Kal2014} and ionic fluids \cite{HolPatPat16, HolPatPat17}.
A generalization of SPT theory for the description of a hard spherocylinder fluid in disordered porous media is presented in \cite{Holovko2014,Holovko2015}. Using the obtained results in our separate paper we are going to consider the generalization of SPT theory for the mixture of hard spheres and hard spherocylinders in random porous media.

\section*{Acknowledgements} 

This project has received funding from the European Unions Horizon 2020 
research and innovation programme under the Marie Sk{\l}odowska-Curie grant agreement 
 No~734276,  and from the State Fund For Fundamental Research (project~N~F73/26-2017).

\ukrainianpart

\title{Ізотропно-нематичний перехід в суміші твердих сфер та твердих сфероциліндрів: застосування теорії масштабної частинки}

\author{М.Ф. Головко, М.В. Гвоздь }
\address{Інститут фізики конденсованих систем НАН України, вул. Свєнціцького, 1, 79011 Львів, Україна}

\makeukrtitle

\begin{abstract}
\tolerance=3000%
Для опису термодинамічних властивостей суміші твердих сфер та твердих сфероциліндрів  розвинуто теорію масштабної частинки. Отримано аналітичні вирази для вільної енергії, тиску та хімічних потенціалів. Мінімізацією вільної енергії сформульовано нелінійне інтегральне рівняння для орієнтаційної унарної функції розподілу. З біфуркаційного аналізу цього рівняння досліджено ізотропно-нематичний фазовий перехід в даній суміші. Показано, що при збільшенні концентрації твердих сфер загальний коефіцієнт упаковки суміші на границях фаз злегка зростає. Представлено порівняння отриманих результатів з даними комп'ютерного моделювання.
\keywords суміш твердих сфер та твердих сфероциліндрів,  ізотропно-нематичний перехід, метод  масштабної частинки

\end{abstract}


\begin{thebibliography}{99}

\bibitem{Vroege1992} Vroege G.J., Lekkerkerker H.N.W., Rep. Prog. Phys., 1992, {\bf55}, 1241,
      \bibdoi{10.1088/0034-4885/55/8/003}.
\bibitem{Franco-Melgar2008} Franco-Melgar M., Haslam A.J., Jackson G., Mol. Phys., 2008, {\bf106}, 649,
      \bibdoi{10.1080/00268970801926958}.
\bibitem{Onsager1949} Onsager L., Ann. N.Y. Acad. Sci., 1949, {\bf51}, 627,
      \bibdoi{10.1111/j.1749-6632.1949.tb27296.x}.
\bibitem{Koda1996} Koda T., Numajiri M., Ikeda S., J. Phys. Soc. Jpn., 1996, {\bf65}, 3551,
      \bibdoi{10.1143/JPSJ.65.3551}
\bibitem{Cotter1970} Cotter M.A., Martire D.E., J. Chem. Phys., 1970, {\bf 52}, 1909,
      \bibdoi{10.1063/1.1673232}.
\bibitem{Cotter1974} Cotter M.A., Phys. Rev. A, 1974, {\bf 10}, 625,
      \bibdoi{10.1103/PhysRevA.10.625}.
\bibitem{Lasher1970} Lasher G., J. Chem. Phys., 1970, {\bf 53}, 4141,
      \bibdoi{10.1063/1.1673914}.
\bibitem{Cotter1979} Cotter M.A., In: The Molecular Physics of Liquid Crystals, 
    Luckhurst G.R., Gray G.W. (Eds.),  Academic Press, London, 1979, 169--189.
\bibitem{Cotter1978} Cotter M.A., Wacker D.C., Phys. Rev. A, 1978, {\bf18}, 2669,
      \bibdoi{10.1103/PhysRevA.18.2669}.
\bibitem{Parsons1979} Parsons J.D., Phys. Rev. A, 1979, {\bf19}, 1225,
      \bibdoi{10.1103/PhysRevA.19.1225}.
\bibitem{Lee1987} Lee S.D., J. Chem. Phys., 1987, {\bf87}, 4972,
      \bibdoi{10.1063/1.452811}.
\bibitem{Lee1988} Lee S.D., J. Chem. Phys., 1988, {\bf89}, 7036,
      \bibdoi{10.1063/1.455332}.
\bibitem{Holovko2014} Holovko M., Shmotolokha V., Patsahan T., J. Mol. Liq., 2014, {\bf189}, 30,
      \bibdoi{10.1016/j.molliq.2013.05.030}.
\bibitem{Holovko2015} Holovko M., Shmotolokha V., Patsahan T., In: Physics of Liquid Matter: Modern Problems, Bulavin~L., Lebovka~N.~(Eds.), Springer, Heidelberg, 2015, 3--30. 
\bibitem{Galindo2003} Galindo A., Haslam A.J., Varga S., Jackson G., Vanakaras A.G.,
    Photinos D.J., Dunmur D.A., J. Chem. Phys., 2003, {\bf119}, 5216,
      \bibdoi{10.1063/1.1598432}.
\bibitem{Cinacchi2004} Cinacchi G., Mederos L., Velasco E., J. Chem. Phys., 2004, {\bf121}, 3854,
      \bibdoi{10.1063/1.1774153}.
\bibitem{Lago2004} Lago S., Cuetos A., Mart\'inez-Haya B., Rull L.F., J. Mol. Recognit., 2004, {\bf17}, 417,
      \bibdoi{10.1002/jmr.704}.
\bibitem{Vesely2005} Vesely F.J., Mol. Phys., 2005, {\bf103}, 679,
      \bibdoi{10.1080/00268970512331328686}.
\bibitem{Martinez-Raton2006} Martinez-Rat\'on Y., Cinacchi G., Velasco E., Mederos L., Eur. Phys. J. E, 2006, {\bf21}, 175,
    \\  \bibdoi{10.1140/epje/i2006-10058-4}.
\bibitem{Cuetos2007} Cuetos A., Mart\'inez-Haya B., Lago S., Rull L.F., Phys. Rev. E, 2007, {\bf75}, 061701,\\
      \bibdoi{10.1103/PhysRevE.75.061701}.
\bibitem{Cuetos2008} Cuetos A., Galindo A., Jackson G., Phys. Rev. Lett., 2008, {\bf101}, 237802,
      \bibdoi{10.1103/PhysRevLett.101.237802}.
\bibitem{Malijevsky2008} Malijevsk\'y A., Jackson G., Varga S., J. Chem. Phys., 2008, {\bf129}, 144504,
      \bibdoi{10.1063/1.2982501}.
\bibitem{Belli2012} Belli S., Dijkstra M., van Roij R., J. Phys.: Condens. Matter, 2012, {\bf24}, 284128,\\
      \bibdoi{10.1088/0953-8984/24/28/284128}.
\bibitem{Gamez2013} G\'amez F., Acemel R.D., Cuetos A., Mol. Phys., 2013, {\bf111}, 3136,
      \bibdoi{10.1080/00268976.2013.771802}.
\bibitem{Wu2015} Wu L., Malijevsk\'y A., Jackson G., M\"uller E.A., Avenda\~no C., J. Chem. Phys., 2015, {\bf143}, 044906,
      \bibdoi{10.1063/1.4923291}.
\bibitem{Agren1975} {\AA}gren G., Phys. Rev. A, 1975, {\bf11}, 1040,
      \bibdoi{10.1103/PhysRevA.11.1040}.
\bibitem{Holovko2009} Holovko M.,  Dong W., J. Phys. Chem. B, 2009, {\bf 113}, 6360,
      \bibdoi{10.1021/jp809706n}.
\bibitem{Chen2010} Chen W., Dong W., Holovko M., Chen X.S., J. Phys. Chem. B, 2010,
      {\bf 114}, 1225,
      \bibdoi{10.1021/jp9106603}.
\bibitem{Patsahan2011} Patsahan T., Holovko M., Dong W., J. Chem. Phys., 2011, {\bf 134}, 074503,
      \bibdoi{10.1063/1.3532546}.
\bibitem{Reiss1959} Reiss H., Frisch H.L., Lebowitz J.L., J. Chem. Phys., 1959, {\bf 31}, 369,
      \bibdoi{10.1063/1.1730361}.
\bibitem{Reiss1960} Reiss H., Frisch H.L., Helfand E., Lebowitz J.L., J. Chem. Phys., 1960,
      {\bf 32}, 119,
      \bibdoi{10.1063/1.1700883}.
\bibitem{Lebowitz1965} Lebowitz J.L., Helfand E., Praestgaard E., J. Chem. Phys., 1965,
      {\bf 43}, 774,
      \bibdoi{10.1063/1.1696842}.
\bibitem{Holovko2010} Holovko M., Shmotolokha V., Dong W., Condens. Matter Phys., 2010, {\bf 13}, 23607,
     \bibdoi{10.5488/CMP.13.23607}.
\bibitem{Gray1984} Gray C.G., Gubbins K.E., Theory of Molecular Fluids, Claredon Press, Oxford, 1984.
\bibitem{Chen2016} Chen W., Zhao S.L., Holovko M., Chen X.S., Dong W., J. Phys. Chem. B, 2016, {\bf120}, 5491,\\
      \bibdoi{10.1021/acs.jpcb.6b02957}.
\bibitem{Tuinier2007} Tuinier R., Taniguchi T., Wensink H.H., Eur. Phys. J. E, 2007, {\bf 23}, 355,
      \bibdoi{10.1140/epje/i2007-10197-0}.
\bibitem{Kayser1978} Kayser R.F. (Jr.), Ravech\'e H.J., Phys. Rev. A, 1978, {\bf 17}, 2067,
      \bibdoi{10.1103/PhysRevA.17.2067}.
\bibitem{Herzfeld1984} Herzfeld J., Berger A.E., Wingate J.W., Macromolecules, 1984, {\bf 17}, 1718,
      \bibdoi{10.1021/ma00139a014}.
\bibitem{Lekkerkerker1984} Lekkerkerker H.N.W., Coulon Ph.,  Van Der Haegen R.,
  Deblieck R., J. Chem. Phys., 1984, {\bf 80}, 3427,
      \bibdoi{10.1063/1.447098}.
\bibitem{Chen1993} Chen Z.Y., Macromolecules, 1993, {\bf 26}, 3419,
      \bibdoi{10.1021/ma00065a027}.
\bibitem{Bolhuis1997} Bolhuis P., Frenkel D., J. Chem. Phys., 1997, {\bf 106}, 666,
      \bibdoi{10.1063/1.473404}.
\bibitem{HolPat12} Holovko M., Patsahan T.,  Dong W., Condens. Matter. Phys., 2012, {\bf 15}, 23607,
        \bibdoi{10.5488/CMP.15.23607}.
\bibitem{Holovko2013} Holovko M., Patsahan T., Dong W., Pure  Appl. Chem., 2012, {\bf 85}, 115,
      \bibdoi{10.1351/PAC-CON-12-05-06}.
\bibitem{Kal2014} Kalyuzhnyi Y.V., Holovko M., Patsahan T., Cummings P.T., J. Phys. Chem. Lett., 2014, {\bf 5}, 4260,
        \bibdoi{10.1021/jz502135f}.
\bibitem{HolPatPat16} Holovko M.F., Patsahan O.,  Patsahan T.,  J. Phys.: Condens.  Matter, 2016, {\bf 28}, 414003,\\
        \bibdoi{10.1088/0953-8984/28/41/414003}.
\bibitem{HolPatPat17} Holovko M.F.,  Patsahan T.M.,  Patsahan O.V., J. Mol. Liq., 2017, {\bf 235}, 53,
        \bibdoi{10.1016/j.molliq.2016.11.030}.


\end{thebibliography}
\end{document}